\documentclass{article}
\usepackage{spconf,amsmath,graphicx}
\usepackage{tikz}
\usepackage{booktabs}
\usepackage{comment} 
\usepackage{amssymb}

\usepackage{hyperref}

\title{Identifiability of Rotating Stellar Surfaces from Astrometric Jitter}
\name{Jamila S. Taaki$^{1}$ \qquad Lia Corrales$^{2}$ \qquad Alfred O. Hero III$^{3}$}
\address{%
$^{1}$ Michigan Institute for Data Science, University of Michigan, Ann Arbor, USA \\
$^{2}$ Department of Astronomy, University of Michigan, West Hall, Ann Arbor, USA \\
$^{3}$ Department of Electrical Engineering and Computer Science, University of Michigan, Ann Arbor, USA \\
\texttt{tjamila@umich.edu}
}

\begin{document}

\maketitle
\begin{abstract}
Astrometry measures shifts in a star's photocentre and can be used to detect reflex motion due to orbiting exoplanets. Brightness asymmetries (e.g. starspots) rotating in and out of view can also cause apparent motion of the photocenter, termed astrometric jitter, that has previously been considered a source of noise. Here we explore whether it can be used to map stellar surfaces. We derive a Cramér-Rao bound on the minimum variance for which a stellar surface can theoretically be estimated, quantifying the information content in rotational astrometric jitter. To regularize and break singularities in the Fisher information, we impose a spatial-smoothness Gaussian–Markov random field prior. A key challenge in mapping surfaces arises for stars with unknown rotational axis inclinations, requiring joint estimation of the inclination and the stellar surface. We characterize the coupling between them and quantify the precision gain when inclination is known versus unknown. 
\end{abstract}
\section{INTRODUCTION}
\label{sec:intro}
Astrometry can be used to measure the gravitational reflex motion of a star due to an orbiting companion exoplanet, or star, via precise calibration of the position of the stellar point-spread-function on a CCD sensor \cite{perry_book}. Astrometry can be used to both detect companion objects, as well as estimate their masses. Among techniques to detect Earth-mass planets in the habitable zones of stars, including radial velocity and direct imaging \cite{exop}, astrometry is anticipated to be uniquely powerful as it is sensitive to planets with long orbital periods \cite{shao}. As a star rotates, dark starspots or bright regions of the photosphere may come in and out of view, changing the measured photo-center\cite{sun-jitter} \cite{meunier}. A starspot of size $3 \%$ the radius of the star, can yield an astrometric signal of the same scale as that of an Earth-analog and active regions occupying $> 1 \%$ of the visible disc are common on stars \cite{strass}. Using the total brightness of a star measured throughout a rotation has been widely explored as a method to map stellar surfaces \cite{russell,cowan,Luger_2021} and applied in the context of high-precision photometry from \emph{Kepler} and \emph{TESS} telescopes. Building on \cite{morris2018spotting}, who propose that upcoming astrometric data (expected 2026) from the Gaia telescope \cite{gaia} could be sensitive to starspot induced photocenter jitter, in concurrent work (Taaki et al., ApJ in prep.) we formulate a forward model that uses this rotational signal (the first image moments) to resolve stellar surface features. Stellar activity and starspots are a major contaminant in exoplanet measurements and limit the achievable sensitivity  \cite{limit} \cite{pont}, \cite{rackham}. Spatially resolving the surfaces of stars adds information that can be used to identify true exoplanetary signals, as well as to understand the magnetic activity that leads to the formation of starspots.  
\begin{figure}[htb]
\begin{minipage}[b]{1.0\linewidth}
  \centering
  \centerline{\includegraphics[width=8.5cm]{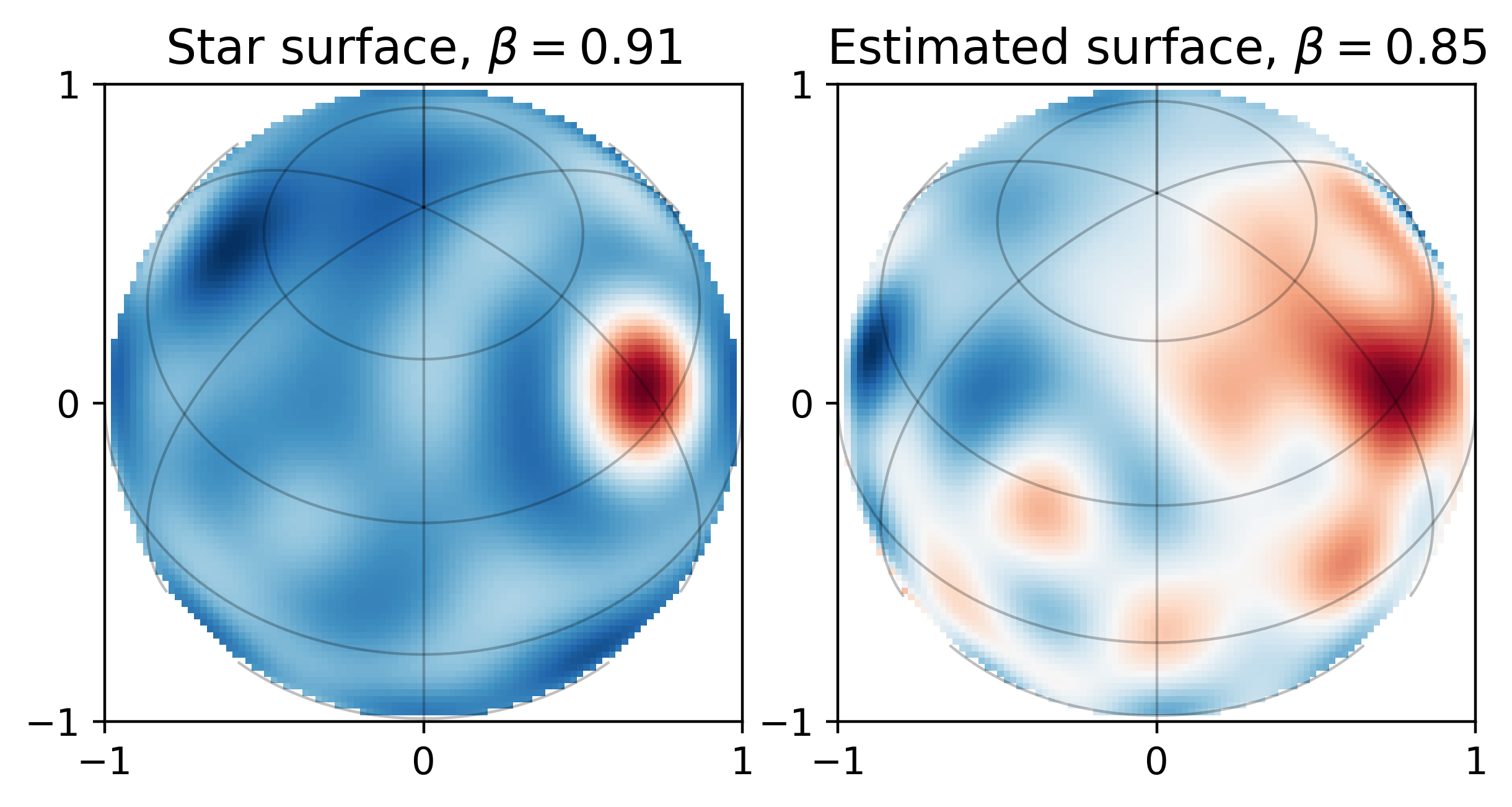}}
\end{minipage}
\caption{Simulated star (left), recovered inclination and stellar surface (right) as the MAP estimate for our measurement model. $L = 9$ for both. The simulated noise level $\sigma^2$ is at $5 \%$ of the signal standard deviation. }
\label{fig:reco}
\end{figure}

However, stellar surface estimation is difficult for several reasons. First, the stellar surface is infinite-dimensional. To address this we use spherical harmonics to represent the stellar surface enabling a sensitivity-ordered truncation of the representation based on the instrument resolution. Second, stars rotate with an often-unknown axis inclination. In this work, we quantify identifiability of the surface intensity via Cramér Rao bounds (CRBs), \cite{Cramer1946}\cite{Rao1992}, particularly when the inclination is deterministic but unknown and must be jointly estimated. An example reconstruction using the proposed measurement model in Section \ref{sec:forward} is shown in Figure \ref{fig:reco}. We derive and analyse the hybrid CRB \cite{rockah} \cite{messer} under this forward model in which measurements are linear in the surface but nonlinearly parameterised by unknown inclination, and we explicitly characterise how information couples between the surface and inclination. 
We further show how a Gaussian Markov random field (GMRF) prior can be gainfully applied as a model of a stellar surface to perform joint estimation of inclination and surface intensity. Similar isotropic scale dependent priors have been used for mapping exoplanets in \cite{farr} and more generally in spatial modeling \cite{gmrf_spatial}. Improved stellar surface reconstructions can help mitigate stellar activity-induced bias and boost sensitivity, bringing us closer to finding and characterizing small Earth-sized exoplanets.

\section{FORWARD MODEL} \label{sec:print}

\begin{figure}[htb]

\begin{minipage}[b]{1.0\linewidth}
  \centering
  \centerline{\includegraphics[width=8.5cm]{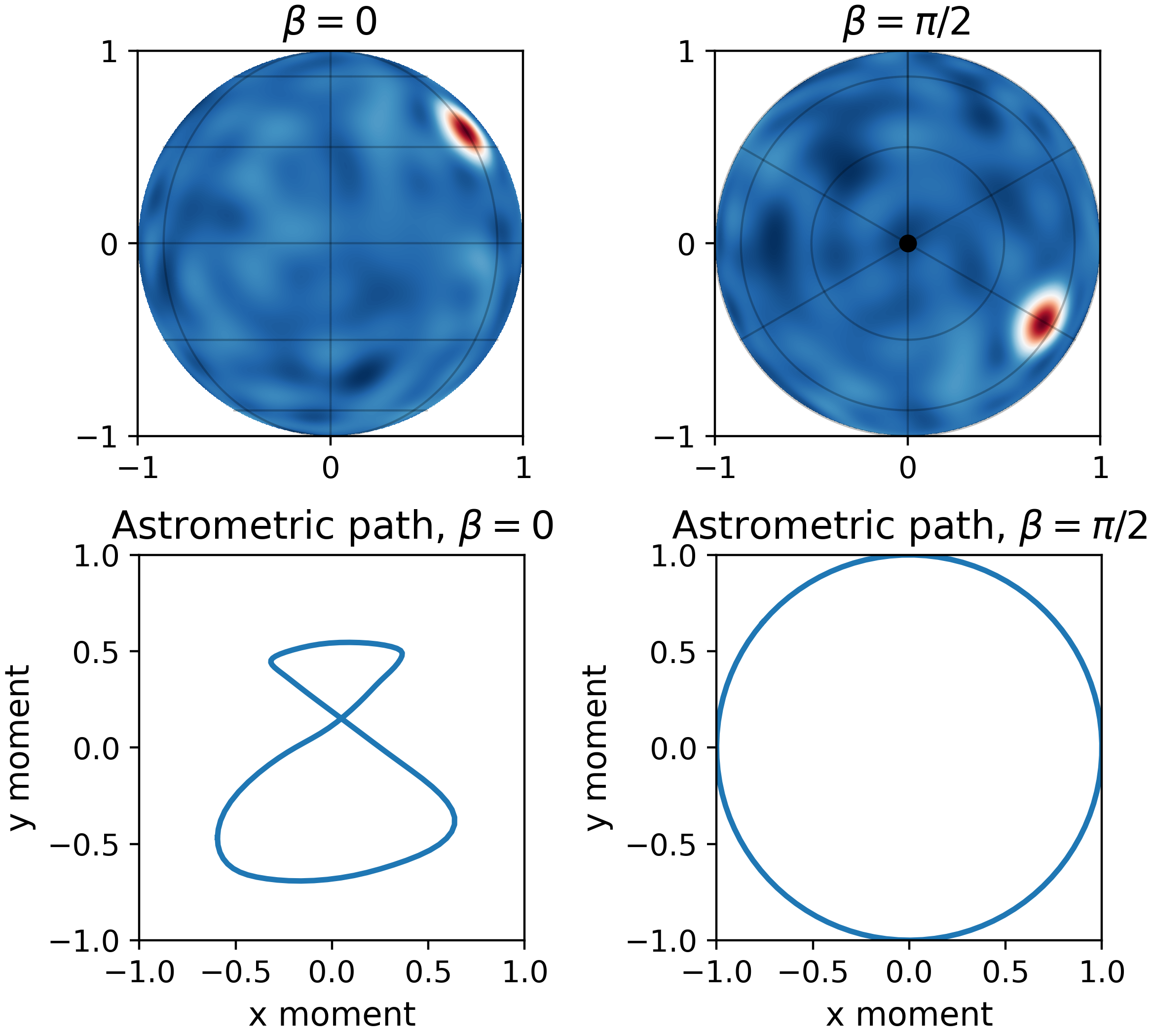}}
\end{minipage}
\caption{Simulated star and astrometric moments at two inclinations (face-on $\beta = 0$ and polar $\beta = \pi/2$). The first row shows a simulated stellar surface as a GMRF, with a starspot from the observers perspective. The surface is represented by $L=20$ spherical harmonics. The second row shows the noiseless astrometric signal observed.}
\label{fig:res}
\end{figure}

Starting with the forward model, we expand the (static) stellar surface in spherical harmonics and express the astrometric first moments (photo-center motion from the observers perspective) over a full stellar rotation as a linear mapping. The linear mapping however depends non-linearly on the inclination of the stars rotational axis inclination (relative to the observer). An example is shown in Figure \ref{fig:res}.
By reparameterizing the measurement model in a separable form over time and inclination, we are able to cleanly derive a CRB below in Section \ref{sec: crb}.

\subsection{Measurement model} \label{sec:forward}
The stars unknown surface intensity is described by a vector $\mathbf{s} \in \mathcal{C}^{(L+1)^2}$ of spherical harmonic coefficients up to degree $L$. We parameterize the stars rotation by the inclination $\beta \in[0,\pi/2]$ and a spin rate $\omega$ (assumed known). The measurement matrix $A(\beta)$ that acts on $\mathbf{s}$ encapsulates the time-dependent rotation, as well as the mapping from the visible surface $\mathbf{s}$ to photocentre measurements along the observer axes $x_{\rm obs}$ and $y_{\rm obs}$. 

Taking $N\!\ge\!2L{+}1$ measurement times and stacking the photocentre first-moment measurements along the observer axes as
$\mathbf{y}=\big[\,\mathbf{y}_{x_{\rm obs}}^T,\ \mathbf{y}_{y_{\rm obs}}^T\,\big]^T \in \mathbb{C}^{2N}$, the measurements $\mathbf{y}$ are linear in  $\mathbf{s}$:
\begin{align}\label{eq:model}
\boxed{\;
\mathbf{y} \;=\; A\!\big(\beta\big)\,\mathbf{s}\;+\;\mathbf{n},
\qquad A\!\big(\beta\big)\in\mathbb{C}^{2N\times (L+1)^2},
\;}
\end{align}
With i.i.d.\ Gaussian noise $\mathbf{n}\sim\mathcal{N}(\mathbf{0},\sigma^2 I)$.
\subsection{Signal Model}
\underline{Stellar surface: } We use the complex spherical harmonics defined for spherical coordinates $\theta \in [-\pi/2, \pi/2]$ and $\phi \in [0, 2\pi]$ as a basis for the stellar surface. These are given by:
\begin{align}
Y_\ell^{m}(\theta,\phi)
= N_{\ell}^m \,P_\ell^{m}(\cos\theta)\,e^{i m \phi},
\end{align}
indexed by $\ell=0,1,\dots,L,\ \ m=-\ell,\dots,\ell$, where $P_l^m$ are the Legendre polynomials and $N_l^m$ is a constant. Higher order terms are decreasing in spatial scale and orthonormal on $S^2$. Let the spherical stellar surface be expanded in this basis up to degree \(L\) with coefficients
\(\mathbf{s} \in \mathbb{C}^{(L+1)^2}\). 
We retain the complex spherical-harmonic basis for algebraic convenience, but the physical surface is modeled as real, enforcing the constraint: $\mathbf{s}$ by \( s_{\ell,-m}=(-1)^m\,s_{\ell m}^{*},\; s_{\ell 0}\in\mathbb{R}.\) This constraint reduces the effective degrees of freedom by half (for \(m>0\)). 
\noindent \underline{Astrometry signal: } Let \(\mathbf{k}_h^\ell\in\mathbb{C}^{2\ell+1} : h\in\{x_{\rm obs},y_{\rm obs}\}\) be the first-moments (astrometric kernel) of a static \(Y_\ell^m\) from the $x_{obs}$ and $y_{obs}$ axis in the observers frame and for a static \(Y_\ell^m\). For an observer measuring the 2D projection along the 3D y-axis, these have the form:

\begin{align}
k_x^{\ell,m}&=
\begin{cases}
\displaystyle \!\int_{\Omega_{\rm vis}} x_{\rm obs}\,V\,Y_\ell^{m}\,d\Omega,
& \ell\ \text{odd}\ \wedge\ m\ \text{odd}, \\
0,& \text{otherwise},
\end{cases}\\[-2pt]
k_y^{\ell,m}&=
\begin{cases}
\displaystyle \int_{\Omega_{\rm vis}} y_{\rm obs}\,V\,Y_\ell^{m}\,d\Omega,
& \ell\ \text{odd}\ \wedge\ m\ \text{even}, \\
0,& \text{otherwise,}
\end{cases}
\end{align}

where $V$ encodes visibility/foreshortening and $d\Omega$ is the differential solid angle. 
We model the stellar rotation with Wigner–$D$ rotation matrices $D^\ell (R)$ acting on each $\mathbf{k}_h^\ell$ vector.  
The first-moments of the star $\mathbf{s}$ after a rotation $R^{-1}$ are $\mu_h(R) =\;\sum_{\ell =0}^L\big\langle \mathbf{s}^\ell,\ D^\ell(R) \mathbf{k}_h^\ell\big\rangle$.

We obtain separable rotations over time and inclination. Inclination mixes orders \(m \in \{ -\ell \dots \ell \} \) within each \(\ell\) as $\mathbf{k}_h^{'\ell} = d^\ell_\beta \mathbf{k}_h^\ell $, where $d^\ell_\beta$ are the small Wigner-$D$ terms. Collecting by degree gives the block–row matrices:

\begin{align}
B_\beta^h = \big[\,\mathrm{diag}(d^0_\beta\mathbf{k}_h^0) \cdots \;\; \mathrm{diag}(d^L_\beta\mathbf{k}_h^L)\,\big],
\end{align} with shape $B_\beta^h \in \mathbb{C}^{(2L+1)\times(L+1)^2}$.

The matrix \(W_\omega \in \mathbb{C}^{N\times(2L+1)}\) for known rotation rate \(\omega\) and measurement times \(\{t_k\}_{k=1}^N\) has components $[W_\omega]_{k,m} \;=\; e^{-i m \omega t_k}$.

We construct the separable signal model as:
\begin{align}
\mathbf{y} = A(\beta)\,\mathbf{s} + \mathbf{n}, 
\qquad 
A(\beta) \;=\; 
\begin{bmatrix}
W_\omega & \\ & W_\omega
\end{bmatrix}
\begin{bmatrix}
B_\beta^x \\[2pt] B_\beta^y
\end{bmatrix}
\,.
\end{align}
The stacked matrix $B_\beta$ is wide and has rank at most $4L$. For all $\beta \in [0, \pi/2]$ the null-space of $A(\beta)$ is therefore non-empty and so the surface is not uniquely constrained even if $\beta$ is known. The zeroth harmonic mode is always in the null-space of the forward model, as are all spherical harmonics with even $\ell > 2$. 

\section{Cramér Rao Bound}\label{sec:page}
\subsection{Bayesian Likelihood} \label{sec: crb}
We treat both surface $\mathbf{s}$ and inclination $\beta \in [0,\pi/2],$ as unknown. Without prior information, the surface and inclination are jointly unidentifiable. Therefore we apply a Gaussian prior on the surface coefficients \(\mathbf{s} \sim \mathcal{CN}(\boldsymbol{\mu}, \Sigma)\),
where \(\Sigma\) parameterises a Gaussian–Markov random field. Under \eqref{eq:model} the joint negative log-density is
\begin{equation}\label{eq:ll-generic}
-\log p(\mathbf{y},\mathbf{s}\mid \beta)\ \propto\
\frac{1}{\sigma^2}\,\big\|\mathbf{y} -A(\beta)\mathbf{s}\big\|_2^2
\;+\; (\mathbf{s}-\boldsymbol{\mu})^{H}\Sigma^{-1} (\mathbf{s}-\boldsymbol{\mu}) .
\end{equation}

\subsection{Hybrid Fisher Information}
We form the hybrid Fisher information \cite{messer} for mixtures of deterministic/random parameters, related to the Bayesian CRM \cite{vantrees}. Although less tight than the CRB, the hybrid CRB can be evaluated in this setting and reveals key aspects of system behaviour.

Let \(\boldsymbol{\theta}=[\,\beta,\mathbf{s}^T\,]^T\) and \(\mathbf{m}(\boldsymbol{\theta})=A(\beta)\mathbf{s}\) where \(p(\mathbf{y}|\boldsymbol{\theta})=\mathcal{N}(\mathbf{m}(\boldsymbol{\theta}),\sigma^2 I)\). By applying the Slepian–Bangs formula the hybrid CRB can be stated as:
\[
\big[\mathcal{I}(\boldsymbol{\theta})\big]_{ij}
=\mathbb{E}_{\mathbf{s}}
\!\left[\frac{1}{\sigma^2}\,
\frac{\partial \mathbf{m}^{H}}{\partial \theta_i}\,
\frac{\partial \mathbf{m}}{\partial \theta_j}\right]
+
\mathbb{E}_{\mathbf{s}}\!\left[
\frac{\partial \log p(\mathbf{s})}{\partial \theta_i}\,
\frac{\partial \log p(\mathbf{s})}{\partial \theta_j}
\right],
\]
The hybrid Fisher information blocks are therefore:
\begin{equation}\label{eq:Jhyb_bar}
\mathcal{I}=
\begin{bmatrix}
{\mathcal{I}}_{\beta\beta} & {\mathcal{I}}_{\beta s}\\[2pt]
{\mathcal{I}}_{s\beta} & {\mathcal{I}}_{ss}
\end{bmatrix}.
\end{equation}

With \(A(\beta)=W_\omega B_\beta\), $\frac{d A(\beta)}{d\beta} = W_\omega \frac{d B_\beta}{\beta}$ and \(W_\omega^{H}W_\omega=N I\). Define \(B'_\beta=\frac{d B_\beta}{d\beta}\). Then:
\begin{align}
{\mathcal{I}}_{\beta\beta}
&= \frac{N}{\sigma^2}\!\left(\operatorname{tr}(B_\beta'^H B_\beta' \Sigma)+\boldsymbol{\mu}^H B_\beta'^H B_\beta' \boldsymbol{\mu}\right),\\
{\mathcal{I}}_{\beta s} \;=\; {\mathcal{I}}_{s\beta}^H
&= \frac{N}{\sigma^2}\,\boldsymbol{\mu}^H B_\beta'^H B_\beta,\\
{\mathcal{I}}_{ss}
&= \frac{N}{\sigma^2}\,B_\beta^H B_\beta + \Sigma^{-1}.
\end{align}
\begin{align}
B'_\beta &= \big[\,\mathrm{diag}(\tfrac{d d_\beta^{0}}{d\beta}\,\mathbf{k}^{0})\ \cdots\
\mathrm{diag}(\tfrac{d d_\beta^{L}}{d\beta}\,\mathbf{k}^{L})\,\big],\\
\frac{d d^\ell_\beta}{d \beta} &= -\,i\, D^\ell\!\left(\tfrac{\pi}{2}, \tfrac{\pi}{2}, 0\right)
\,\mathrm{diag}(-\ell,\ldots,\ell)\,
D^\ell\!\left(\beta+\pi, \tfrac{\pi}{2}, \tfrac{\pi}{2}\right),
\end{align}
where $D^\ell$ are Wigner $D$ matrices. The Fisher information is singular.

\subsection{Stellar Surface CRB}

The Fisher information for the surface coefficients \(\mathbf{s}\) obtained from the Schur complement is:
\begin{align} \
\mathcal{I}_{\mathbf{s}\mid\beta}
= \mathcal{I}_{ss}- \mathcal{I}_{s\beta}\,\mathcal{I}_{\beta\beta}^{-1}\,\mathcal{I}_{\beta s}\\
= \frac{N}{\sigma^2}\,B_\beta^{H}\!\left(I-\Pi_{B'_\beta\boldsymbol{\mu},\Sigma}\right)\!B_\beta \;+\; \Sigma^{-1}.
\end{align} \label{eq: info_surface} 
Where:
\begin{align}
\Pi_{B'_\beta\boldsymbol{\mu},\Sigma}
=
\frac{ B'_\beta \boldsymbol{\mu}\,\boldsymbol{\mu}^{H} B_\beta'^{H} }
{ \boldsymbol{\mu}^{H} B_\beta'^{H} B_\beta' \boldsymbol{\mu} + \operatorname{tr}(B_\beta'^{H}B_\beta' \Sigma) },
\end{align}
Hence the hybrid CRB for an unbiased estimator of the surface $\mathbf{s}$ is:
\begin{align} \label{eq:crb_s}
\mathrm{Cov}(\widehat{\mathbf{s}})\ \succeq\ \big(\mathcal{I}_{\mathbf{s}\mid\beta}\big)^{-1}.
\end{align}
Because the spherical-harmonic basis is orthonormal, this CRB on the spherical harmonic coefficients translates directly into a bound on surface-estimation precision.

\subsection{Inclination Angle CRB}
Similarly we obtain the Fisher information for the inclination angle $\beta$ as:
\begin{align}
\mathcal{I}_{\beta\mid\mathbf{s}} = {\mathcal{I}}_{\beta\beta} \;-\; {\mathcal{I}}_{\beta s}\,({\mathcal{I}}_{ss})^{-1}\,{\mathcal{I}}_{s\beta} \\
= \frac{N}{\sigma^2}\left[
\boldsymbol{\mu}^H B_\beta'^H \Pi^{\perp}_{B_\beta,\Sigma} B_\beta' \boldsymbol{\mu}
\;+\;
\operatorname{tr}\!\left(B_\beta'^H \Pi^{\perp}_{B_\beta,\Sigma} B_\beta' \Sigma\right) \right]
\label{eq:Ibeta_bar_raw}
\end{align}
With the regularized projector:
\begin{align}
\Pi^{\perp}_{B_\beta,\Sigma}=I-\Pi^{}_{B_\beta,\Sigma}, \\
\Pi^{}_{B_\beta,\Sigma}
=
B_\beta\!\left(B_\beta^{H} B_\beta + \frac{\sigma^2}{N}\,\Sigma^{-1}\right)^{\!-1}\!B_\beta^{H}, 
\end{align}
For an unbiased estimator of the inclination $\beta$, the hybrid CRB is:
\begin{equation}\label{eq:hyb_bounds}
\mathrm{Var}\!\left(\widehat{\beta}\right)\;\ge\;\big(\mathcal{I}_{\beta|\mathbf{s}}\big)^{-1}.
\end{equation}
\subsection{GMRF prior}
We place a Gaussian-Markov Random Field (GMRF) prior on the surface where coefficients with the same degree \(\ell\) share a common weight $\mathbf{s}^\ell\sim\mathcal{CN}\!\big(\boldsymbol{\mu}, q_\ell I_{2\ell+1} \big),$. By the orthonormal spherical harmonic transform, this implies a localized difference penalty in spatial coordinates enforcing smoothness.
This is appropriate since the astrometric kernel weights $\mathbf{k}^\ell$ decay with $\ell$ and components have lower SNR. A scale-dependent prior downweights those modes, improving conditioning.
With $q_\ell \propto (1/ \lambda)(\gamma/\ell)^{\alpha}$, $\gamma$ is a scale parameter, for $\alpha > 1$. Here $\lambda > 0$ a regularisation weighting.

\section{Experiment}
\label{sec:exp-inclination}
In some scenarios, the inclination angle $\beta$ of the rotational axis may be directly measured through independent observations. To evaluate the coupling between $\beta$ and the stellar surface coefficients $\mathbf{s}$, we quantify the impact of knowing $\beta$ on the recovery of $\mathbf{s}$ by comparing the CRB for $\mathbf{s} \mid \beta$ (when $\beta$ is unknown) in Equation \ref{eq:crb_s} with the CRB when $\beta$ is known (simply given by $\mathcal{I}_{ss}^{-1}$). 

The maximal loss in precision in Equation \ref{eq: info_surface} is found when $B'_\beta \boldsymbol{\mu}$ is aligned with the largest singular vector of $B_\beta$, we set $\mathbf{s}$ to achieve this worst case.

We use the trace of the CRB as a comparison metric and report the relative gain (in \%) from knowing \(\beta\):
\begin{align}
\label{eq:gain}
\text{gain} \;=\;
\frac{\operatorname{tr}\big(\mathcal{I}_{\mathbf{s} \mid \beta}^{-1}\big) - \operatorname{tr}\big(\mathcal{I}_{\mathbf{s}\mathbf{s}}^{-1}\big)}
     {\operatorname{tr}\big(\mathcal{I}_{\mathbf{s}\mathbf{s}}^{-1}\big)}.
\end{align}

Experiments sweep \(\beta\in[0,\tfrac{\pi}{2}]\) and:
(a) vary \(\sigma^2\in\{0.1,1,10\} \cdot N\) at fixed \(L=5\);
and (b) vary \(\alpha \in\{0.5,1,2,3\} \) and plot \(\operatorname{tr}\{\mathcal{I}_{\mathbf{s} |\beta}^{-1}\}\) and  \(\operatorname{tr}\{\mathcal{I}_{\mathbf{s} \mathbf{s}}^{-1}\}\) versus \(L\) at a representative inclination \(\beta=\pi/4\). Prior regularisation is set to \(\lambda=10^{-3}\). 

\begin{figure}[htb]

\begin{minipage}[a]{1.0\linewidth}
  \centering
  \centerline{\includegraphics[width=8.5cm]{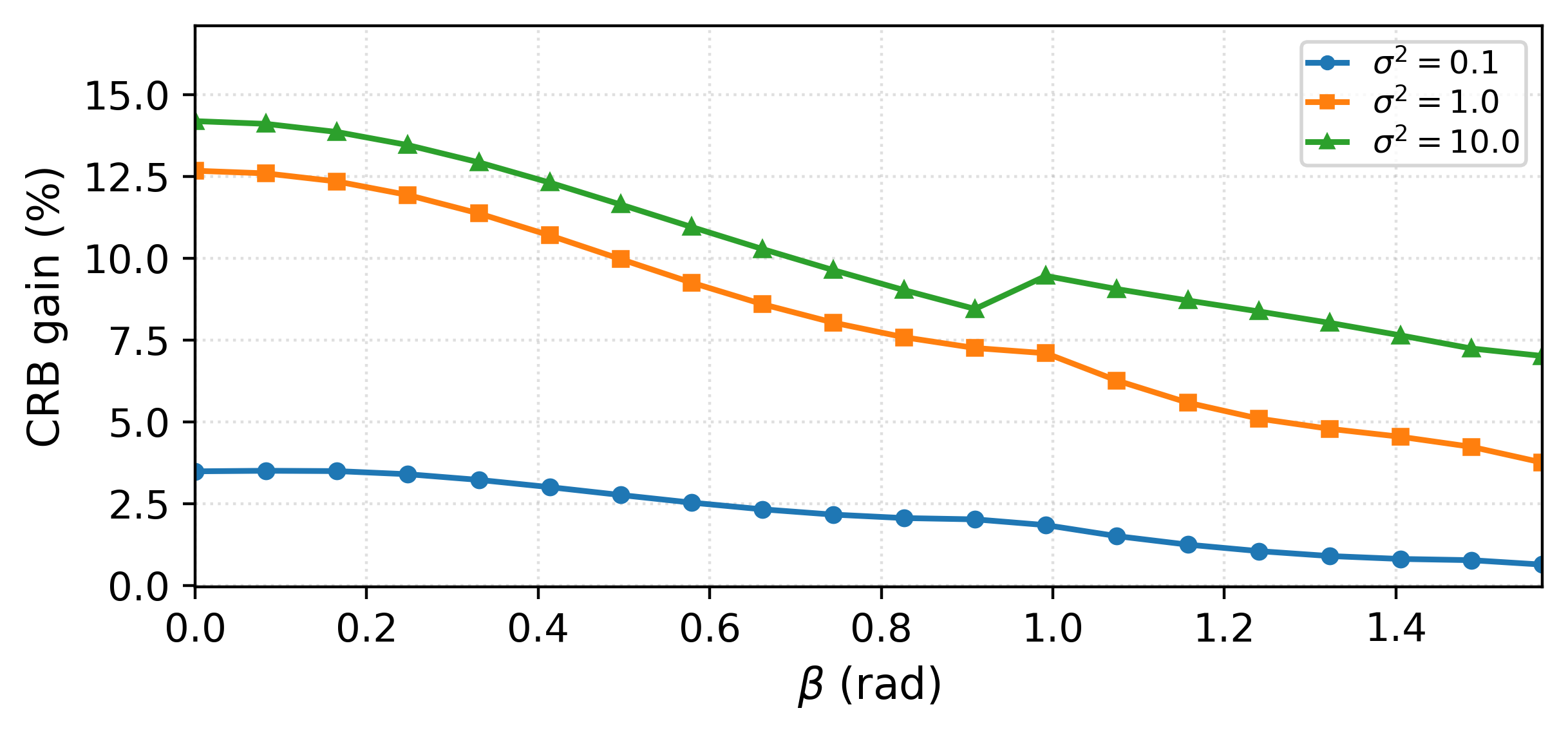}}
  \centerline{(a) }\medskip
\end{minipage}
\begin{minipage}[b]{1.0\linewidth}
  \centering
  \centerline{\includegraphics[width=8.5cm]{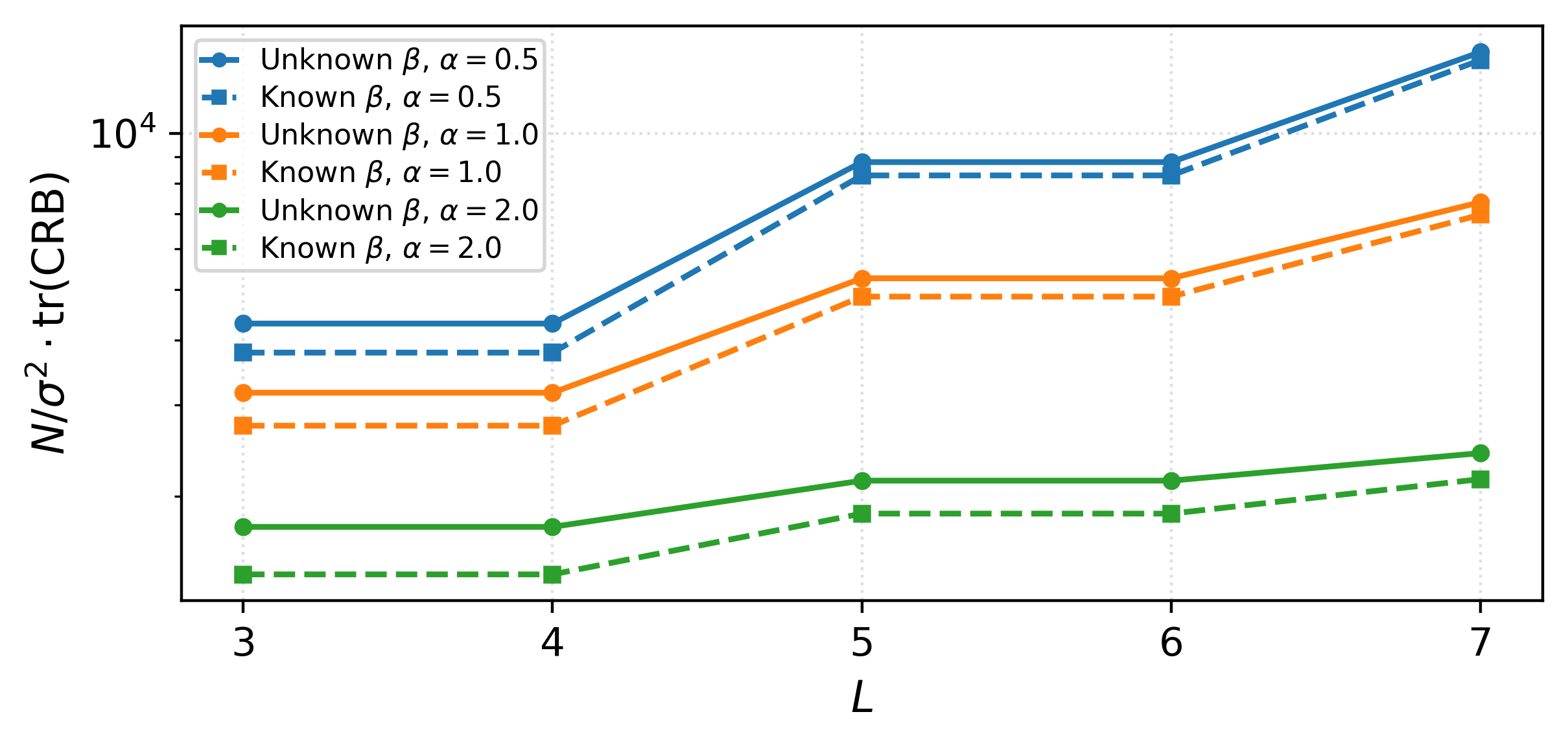}}
  \centerline{(b) }\medskip
\end{minipage}
\caption{The gain in optimal precision in estimating $\mathbf{s}$ from prior knowledge of inclination $\beta$. (a) Gain vs. $\beta$ for $\sigma^2/N\in\{0.1,1,10\}$ at $L=5$; (b) $\operatorname{tr}(\mathrm{CRB})$ vs. $L$ at $\beta=\pi/4$ for the prior exponent factor $\alpha\in\{0.5,1,2,3\}$, for known and unknown $\beta$. Overall, the results underscore that stellar surface mapping from first moments is strongly resolution-limited.}
\label{fig:plots}

\end{figure}

The performance gain is shown in Figure \ref{fig:plots}. Across all parameter ranges the gain is strictly positive. Across inclination $\beta$ the gain is largest at the edge-on perspective and is smallest near the pole–on where $A(\beta)$ is rank deficient and carries minimal information. At fixed $L$ the gain decreases as \(\sigma^2\) increases and the prior dominates both estimators when data are noisy.

Absolute uncertainty curves \(\operatorname{tr}\{\mathrm{CRB}\}\) versus \(L\) grow logarithmically and the relative gain decreases. As $L$ increases the signal model $A(\beta)$ includes more modes with low levels of sensitivity, additionally the maximum rank of the forward model is $4L$ the size of signal space scales as $(L+1)^2$ adding to the information decay. 

\section{Conclusion}\label{sec:rel-conc}
Here we address the joint problem of inferring a rotating stellar surface with inclination from astrometric measurements. We derive hybrid CRBs under the separable model in \eqref{eq:model}. Our results are summarized as follows: \\
(i) Without a prior, interior inclinations lead to zero Fisher information and the pair \((\mathbf{s},\beta)\) is unidentifiable.  \\
(ii) Knowing \(\beta\) strictly improves the surface CRB; the gain is largest for an equatorial observer. However, diminishes as noise increases (Fig.~\ref{fig:plots}a) and as \(L\) grows due to the rank decay of the first-moment operator (\(\le 4L\))].

Astrometric jitter contains recoverable information, but mapping from first moments alone is fundamentally underdetermined and sensitivity-limited. Independent constraints on inclination and physically informed, degree-wise priors are therefore crucial. Guided by these CRBs, future work will develop data-driven priors to improve surface reconstructions and mitigate stellar-activity noise in exoplanet measurements.

\vfill\pagebreak

\bibliographystyle{IEEEbib}
\bibliography{refs}

\end{document}